\documentclass[epj]{svjour}
\usepackage{graphics}
\begin{document}
\title{Validation and verification of MCNP6 against intermediate and 
high-energy experimental data and results by other codes}
\author{S.G. Mashnik
}                     
%
%
\institute{Monte Carlo Codes (XCP-3),
MS: A143,
Los Alamos National Laboratory,
Los Alamos, NM 87545, USA}
%
%
\abstract{
MCNP6, the latest and most advanced LANL transport code representing 
a recent merger of MCNP5 and MCNPX, has been Validated and Verified 
(V\&V) against a variety of intermediate and high-energy experimental 
data and against results by different versions of MCNPX and other codes. 
In the present work, we V\&V MCNP6 using mainly the latest modifications 
of the Cascade-Exciton Model (CEM) and of the Los Alamos version of the 
Quark-Gluon String Model (LAQGSM) event generators CEM03.03 and LAQGSM03.03. 
We found that MCNP6 describes reasonably well various reactions induced by 
particles and nuclei at incident energies from 18 MeV to about 1 TeV per 
nucleon measured on thin and thick targets and agrees very well with 
similar results obtained with MCNPX and calculations by CEM03.03, 
LAQGSM03.03 (03.01), INCL4 + ABLA, and Bertini INC + Dresner evaporation, 
EPAX, ABRABLA, HIPSE, and AMD, used as stand alone codes. Most of several 
computational bugs and more serious physics problems observed in MCNP6/X 
during our V\&V have been fixed; we continue our work to solve all the 
known problems before MCNP6 is distributed to the public.
\PACS{
      {PACS-key}{discribing text of that key}   \and
      {PACS-key}{discribing text of that key}
      } 
} 
\maketitle
\section{Introduction}
\label{intro}
During the past several years, a major effort has been undertaken at the 
Los Alamos National Laboratory (LANL) to develop the transport code 
MCNP6 \cite{1,2}, 
the latest and most advanced Los Alamos transport code representing 
a merger of MCNP5 \cite{3} and MCNPX \cite{4}. 
The work on MCNP6 is not yet completed; 
we continue to solve the observed problems in the current version of MCNP6 
and to develop and improve it further, with a plan to make it available 
officially to the users via RSICC at Oak Ridge, TN, USA, during 2011. 
Before distributing MCNP6 to the public, we must test and validate it on 
as many test-problems as possible, using reliable experimental data. Extensive 
Validation and Verification (V\&V) of our low energy transport code MCNP5 has 
been performed and published for many different test-problems involving 
interactions of neutrons, photons, and electrons with thick and thin targets, 
therefore V\&V of MCNP6 for such problems is important but not very critical. 
On the other hand, our high-energy transport code, MCNPX, was not tested 
against experimental data so extensively, especially for high-energy processes 
induced by protons, and heavy-ions. More important, MCNP6 uses the latest 
modifications of the Cascade-Exciton Model (CEM) and of the Los Alamos version 
of the Quark-Gluon String Model (LAQGSM) event generators CEM03.03 and 
LAQGSM03.03 \cite{5}, and they were not tested extensively in MCNPX. This is why 
it is necessary to V\&V MCNP6 at intermediate and high energies, to test how 
CEM03.03 and LAQGSM03.03 work in MCNP6 and to make sure that the latter properly 
transports energetic particles and nuclei through the matter.

A description of different versions of our CEM and LAQGSM event generators with
 many useful references may be found in our recent lecture \cite{5}. 
Let's us recall 
here only their main assumptions. The basic version of both our CEM and LAQGSM 
event generators is the so-called ``03.01" version, namely CEM03.01 and LAQGSM03.01. 
CEM describes reactions induced by nucleons, pions, and photons at energies 
below $\sim  5$ GeV. LAQGSM describes reactions induced by almost all elementary 
particles as well as by heavy ions at incident energies up to $\sim  1$ TeV/nucleon. 
However, our numerous tests show that CEM provides a little better agreement 
than LAQGSM with experimental data for reactions induced by p, n, $\pi$, 
and $\gamma$ at 
energies below several GeV, therefore we recommend using CEM in MCNP6 to 
describe such reaction at energies below $\sim 3.5$ GeV, and using LAQGSM at 
higher energies and for projectiles not allowed by CEM. 

Both CEM and LAQGSM assume that nuclear reactions occur generally in three 
stages. The first stage is the IntraNuclear Cascade (INC), completely 
different in CEM and LAQGSM, in which primary particles can be re-scattered 
and produce secondary particles several times prior to absorption by, or escape 
from the nucleus. When the cascade stage of a reaction is complete, CEM03.01 
uses the coalescence model to ``create" high-energy d, t, $^3$He, and $^4$He via 
final-state interactions among emitted cascade nucleons, already outside of 
the target. The subsequent relaxation of the nuclear excitation is treated in 
terms of an improved version of the modified exciton model of preequilibrium 
decay followed by the equilibrium evaporation/fission stage of the reaction. 
But if the residual nuclei after the INC have atomic numbers with $A < 13$, CEM 
and
LAQGSM use the Fermi breakup model to calculate their further disintegration 
instead of using the preequilibrium and evaporation models.

The main difference of the following, so-called ``03.02" versions of CEM and 
LAQGSM from the basic ``03.01" versions is that the latter use the Fermi 
breakup model to calculate the disintegration of light nuclei instead of 
using the preequilibrium and evaporation models only after the INC, when 
$A < 13$. It does not use the Fermi breakup model at the preequilibrium, 
evaporation, and fission stages, when, due to emission of preequilibrium 
particles or due to evaporation or to a very asymmetric fission, we get an 
excited nucleus or a fission fragment with $A < 13$. This problem was solved 
in the 03.02 versions of CEM and LAQGSM, where the Fermi breakup model is 
used also during the preequilibrium and evaporation stages of a reaction, 
when we get an excited nucleus with $A < 13$. Finally, the latest, 03.03 versions 
of our codes do not produce any unstable products via very asymmetric fission, 
allowed in very, very rare cases by the previous versions, and have several 
bugs fixed that were observed in the previous versions. More details and 
useful references on CEM and LAQGSM may be found in Ref. \cite{5}.

\section{Validation and verification of MCNP6}
\label{sec:2}

In the present work, we present only a small part of our recent extensive 
V\&V of MCNP6 \cite{6,7} 
using as event-generators mostly CEM03.03 and LAQGSM03.03. 
For convenience, we start with the V\&V of MCNP6 using CEM, followed by results 
using LAQGSM.

Before presenting  our results, let us mention that the 
easiest way to calculate (in MCNP6) spectra of secondary particles
and cross section of products from a thin target  
is to use either the {\bf noact=-2} option on the {\bf LCA} card of the
MCNP6 input file, or the special {\bf GENXS} option of MCNP6. The first option
({\bf noact=-2}) was developed for the MCNPX code and is described in detail
in Section 5.4.6.1 of the MCNPX Manual \cite{18}; it migrated later to
MCNP6 exactly as was done in MCNPX \cite{18}. The second option ({\bf GENXS})
was developed by Dr. Richard Prael especially for MCNP6 and is described
in detail in Ref. \cite{GENXS}. 
Both these documents \cite{18,GENXS} will
be included in the package to be distributed to the MCNP6 users.
The two MCNP6 Primers \cite{6,7}
show examples of MCNP6 input files with both these options. Below,
in cases of test-problems with thin targets,
we show examples of using either the first or the second 
option (or both of them, as shown in Fig. 1).

\subsection{CEM test problems}
\label{sec:2.1}
Let us start our presentation of the V\&V of MCNP6 using CEM with a test-problem 
at a very low energy of 18 MeV, namely the energy spectra of prompt $\gamma$-rays 
from a thick H$_2^{18}$O target bombarded with 18 MeV protons. This problem is of 
interest for positron emission tomography (PET), as $^{18}$F used in PET is usually 
produced via the $^{18}$O(p,n)$^{18}$F reaction and energy spectrum and angular 
distribution 
of neutrons and photons produced in this reaction should be estimated for radiation 
safety and clearance of the production facility \cite{8}. 
Generally, at energies below 
150 MeV, MCNP6 uses data libraries instead of event-generators. But in cases when 
we do not have any data libraries for a particular isotope, MCNP6 must use an 
event-generator, therefore it is necessary to test how well CEM works in MCNP6 
even at such low energies. 

%
\begin{figure*}                                                       
\centering
\resizebox{0.95\hsize}{!}{
\includegraphics{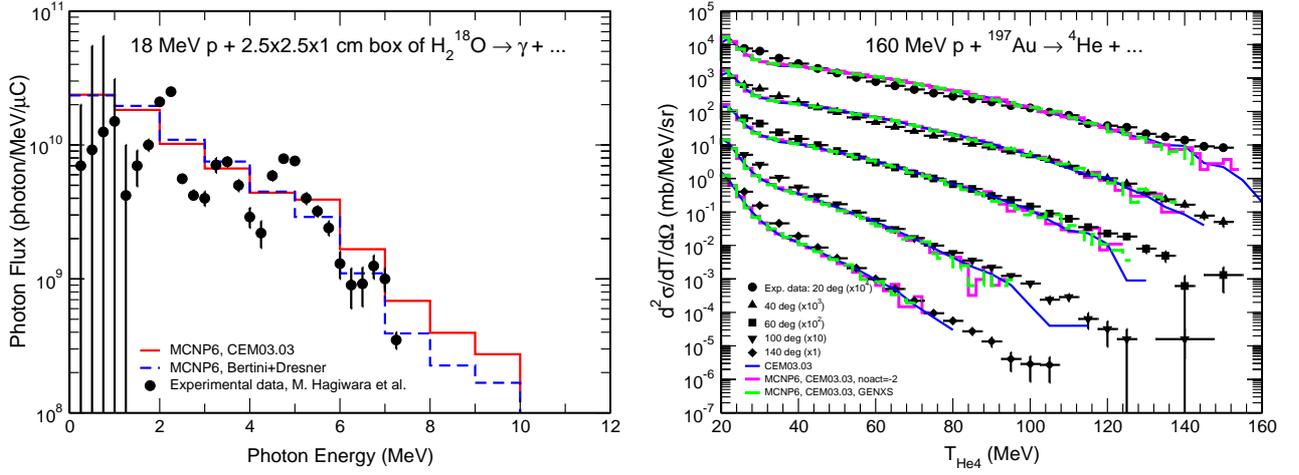}
}
\caption{Comparison of the measured \cite{8} 
energy spectra of prompt $\gamma$-rays 
from a thick H$_2^{18}$O target bombarded with 18 MeV protons with our MCNP6 
results using the CEM03.03 \cite{5} 
and the Bertini INC \cite{9} + Multistage Preequilibrium 
Model (MPM)  \cite{10}
+ Dresner evaporation \cite{11} 
event-generators 
(left plot) and experimental \cite{12} double-differential spectra of $^4$He at 
20, 40, 60, 100, and 140 degrees from interactions of 160 MeV protons with a 
thin $^{197}$Au target compared with calculations by CEM03.03 as presented 
at the recent International Benchmark of Spallation 
Models \cite{13} and with our 
current results by MCNP6 using  CEM03.03 (right plot), as indicated.}
\end{figure*}

%
\begin{figure*}                                          
\vspace*{3mm}       
\centering
\resizebox{0.95\hsize}{!}{
\includegraphics{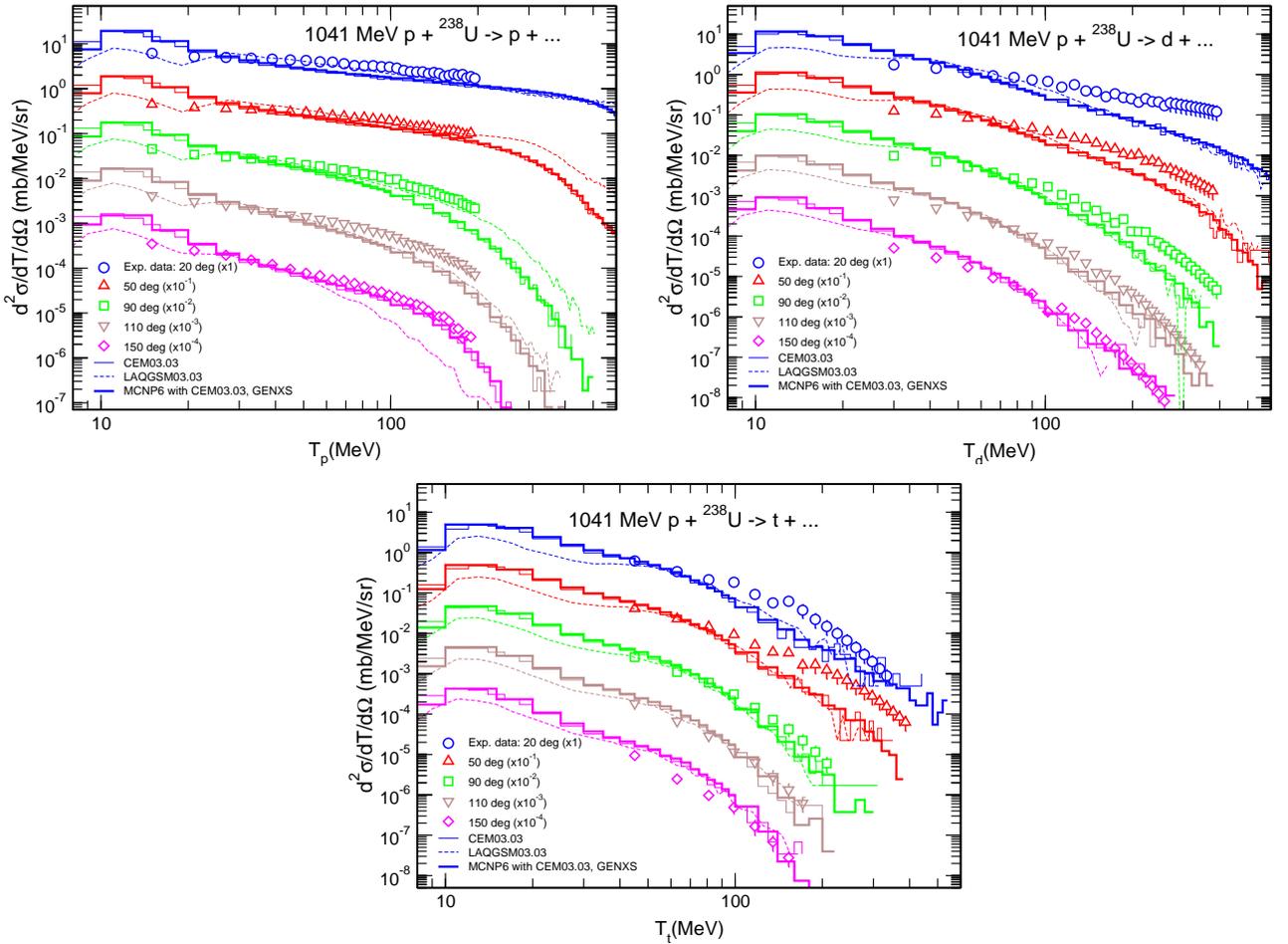}
}
\caption{Experimental \cite{14} 
double-differential spectra of protons, deuterons, 
and tritons at 20$^\circ$, 50$^\circ$, 90$^\circ$, 110$^\circ$, and 150$^\circ$ 
from interactions of 1.041 GeV protons with a thin $^{238}$U target compared 
with calculations by CEM03.03 and LAQGSM03.03 used as stand alone codes and 
with current results by MCNP6 
using the CEM03.03 event-generator, as indicated.}
\end{figure*}

%
\begin{figure*}                                                       
\centering
\resizebox{0.65\hsize}{!}{
\includegraphics{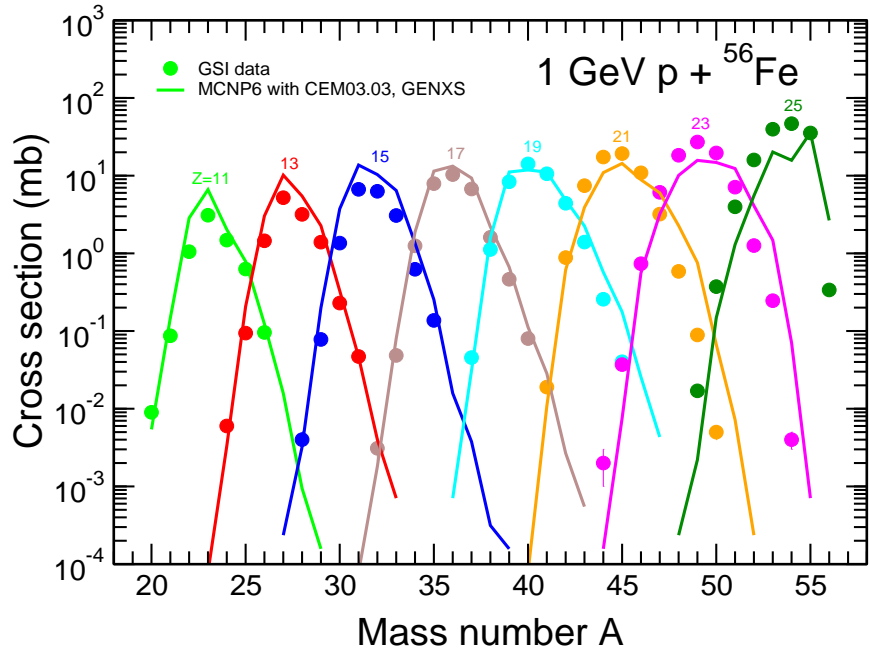}
}
\caption{The measured \cite{15} 
cross sections for the production of elements with 
the charge Z equal to 11, 13, 15, 17, 19, 21, 23, and 25 (filled color circles) 
from the reaction 1 GeV/A $^{56}$Fe + $p$ compared with results MCNP6 using 
the CEM03.03 event-generator, as indicated.}
\end{figure*}

%
\begin{figure*}                                                       
\vspace*{5mm}       
\centering
\resizebox{0.65\hsize}{!}{
\includegraphics{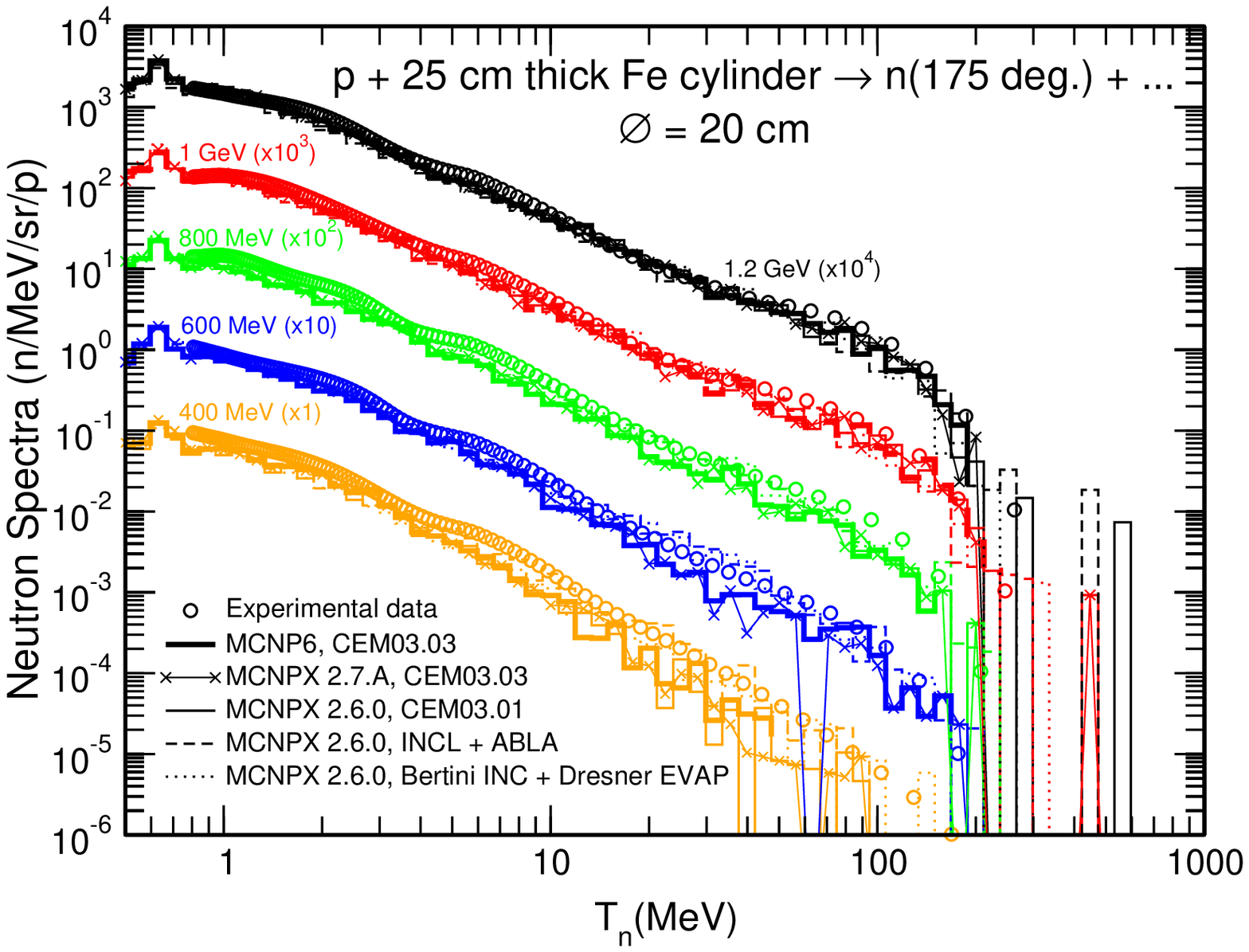}
}
\caption{Experimental \cite{16} neutron spectra at 175 degrees (symbols) 
from a thick Fe cylinder bombarded with 400, 600, 800, 1000, and 
1200 MeV protons compared with results by MCNP6 using the CEM03.03 
event-generator, by MCNPX 2.7.A \cite{17} using the 
CEM03.03 event-generator, and by the 2.6.0 version of MCNPX \cite{18} 
using the CEM03.01, Bertini INC \cite{9} followed by the Multistage 
Preequilibrium Model (MPM) \cite{10} and the evaporation model described 
with the Dresner code EVAP \cite{11}, and by the INCL+ABLA \cite{19,20} 
event-generators, as indicated in legend.}
\end{figure*}

As one can see from the left plot of Fig. 1, MCNP6 with CEM03.03 describes 
well the recently measured \cite{8} 
spectrum of prompt $\gamma$-ray from this reaction 
and agrees reasonably with similar results obtained using the 
Bertini INC \cite{9} + Multistage Preequilibrium 
Model (MPM)  \cite{10}
+ Dresner evaporation \cite{11} 
event-generator.

The right plot in Fig. 1 shows examples of $^4$He spectra emitted from the 160 MeV
$p + ^{197}$Au reaction at five angles as calculated by MCNP6 using CEM03.03 
compared with experimental data \cite{12} and 
with results by CEM03.03 used as a stand 
alone code as presented at the recent International Benchmark of Spallation 
Models \cite{13}. Such reactions are of interest for many applications where gas 
production is important and should be properly accounted; we see that MCNP6 using 
CEM03.03 describes them very well.

Fig. 2 shows examples of $p$, $d$, and $t$ spectra from the 1041 MeV 
$p + ^{238}$U reaction calculated by MCNP6 using CEM03.03 compared with 
experimental data \cite{14} and results by CEM03.03 and LAQGSM03.3 used as 
stand alone codes. Such reactions are of interest for several applications,
especially the cases when gas production must be properly accounted; we see 
that MCNP6 describes them well.

Fig. 3 shows examples of yields of eight elements produced from the reaction 
of 1 GeV/nucleon $^{56}$Fe + $p$ calculated with MCNP6 using CEM03.03 compared 
with recent GSI measurements \cite{15}. Such reactions are of interest for many 
technical applications using iron as a construction material, and we see that 
MCNP6 using CEM03.03 describes them well.

Fig. 4 shows backward angle spectra of neutrons from a thick Fe cylinder 
bombarded by protons of 400, 600, 800, 1000, and 1200 MeV calculated by 
MCNP6 using the CEM03.03  event-generator compared with 
experimental data \cite{16} 
and results by MCNPX 2.7.A \cite{17} using the CEM03.03, 
and by the 2.6.0 version of MCNPX \cite{18} 
using the CEM03.01, Bertini INC \cite{9}
followed by MPM \cite{10} and the Dresner 
evaporation \cite{11}, as well as using the INCL+ABLA \cite{19,20} 
event-generators. 
Such processes are of interest to shielding applications in predicting 
personnel radiation exposure from backward fluxes, to test how the event 
generators work in this ``difficult" kinematics region, and to study the 
mechanisms of cumulative particle production (an academic problem under 
investigation for about four decades but still with many open questions). 
We see that MCNP6 results agree very well with the measured data and 
calculations by other codes. Very similar results were obtained recently 
also for a thick Pb target \cite{21}.

\subsection{LAQGSM test problems}
\label{sec:2.2}

Now, let us present several results on V\&V of MCNP6 using LAQGSM. 
Fig. 5 shows measured \cite{14} 
double-differential spectra of protons at 30$^\circ$, 
70$^\circ$, 90$^\circ$, 110$^\circ$, and 150$^\circ$ from interaction of a 
1042 MeV/nucleon $^{40}$Ar beam with a thin $^{40}$Ca target compared with 
results by MCNP6 using LAQGSM03.03 and by LAQGSM03.03 used as a stand 
alone code. This V\&V problem tests the 
applicability of MCNP6 to calculate production of protons from intermediate 
energy heavy-ion induced reactions for different NASA (shielding for 
missions in space), medical (cancer treatment with heavy-ions), FRIB 
(Facility for Rare Isotope Beams), and for several other
U.S. DOE applications; we see that MCNP6 describes such reactions very well.

%
\begin{figure*}                                                       
\centering
\resizebox{0.65\hsize}{!}{
\includegraphics{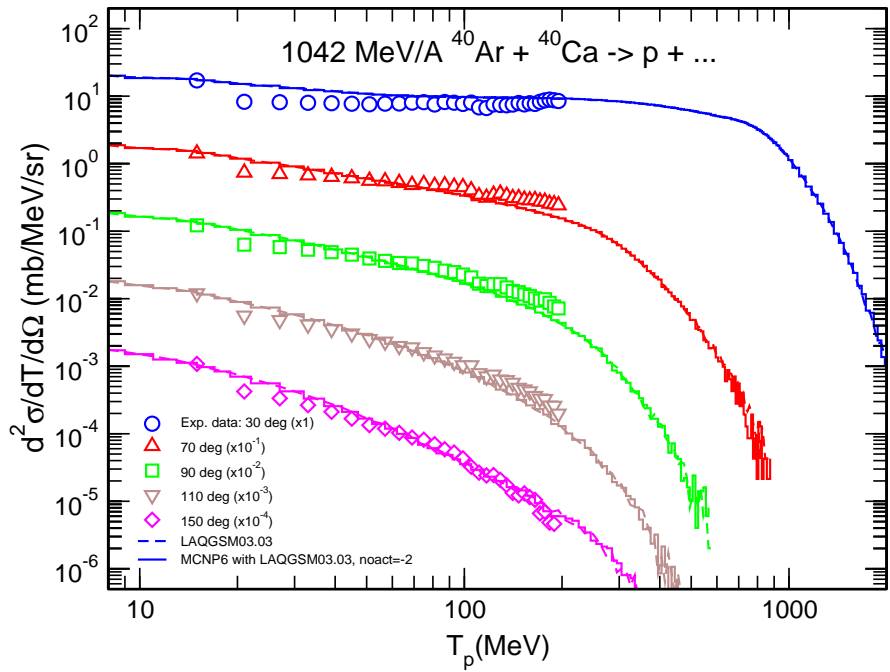}
}
\caption{Experimental \cite{14} proton spectra at 30, 70, 90, 110, and 150 degrees 
(symbols) from a thin $^{40}$Ca target bombarded with a 1042 MeV/nucleon $^{40}$Ar 
beam compared with results by LAQGSM03.03 used as a stand alone code and by MCNP6 
using the LAQGSM03.03 event-generator, as indicated.}
\end{figure*}

In Fig. 6., we test the capability of MCNP6 to describe spectra of complex 
particles from reactions induced by heavy ions at intermediate energies. 
Namely, we compare the experimental \cite{14,22} double-differential spectra of 
$d$, $t$, $^3$He, and $^4$He from thin $^{64}$Cu and $^{238}$U targets 
bombarded with $^{20}$Ne beams of several energies compared with results 
by MCNP6 using the LAQGSM03.03 event-generator and by LAQGSM03.03 used as 
a stand alone code. The interest in such types of reactions is very similar 
to the one listed above regarding Fig. 5. In addition, we like to note that 
light charged particles, i.e., $p$, $d$, $t$, $^3$He, and $^4$He
from any reactions are of a major concern for material damage, as 
helium can cause swelling in structure materials; tritium is often an 
issue from a radioprotection point of view. We see that MCNP6 describes 
such reactions very well.

Fig. 7.  presents experimental \cite{23} yields of the Si ions from the 140 
MeV/nucleon $^{40}$Ca + $^9$Be reaction compared with results by MCNP6  
using the LAQGSM03.03 and by LAQGSM03.03 used as a stand alone code, 
as well as results by EPAX \cite{24}, ABRABLA \cite{25}, 
HIPSE \cite{26}, and AMD \cite{27} 
from \cite{23}. Such reactions are of interest for FRIB (a modification and 
continuation of the former Rare Isotope Accelerator (RIA) project) and 
the measurements \cite{23} were performed especially to support RIA. We see 
that MCNP6 describes these measurements very well and is not worse than 
other models. We obtained similar results for many other reactions 
measured for FRIB/RIA at the National Superconducting Cyclotron 
Laboratory (NSCL) in East Lansing, MI, USA, allowing us to conclude 
that MCNP6 can be a useful and reliable tool for FRIB simulations.

%
\begin{figure*}                                                       
\centering
\resizebox{1.00\hsize}{!}{
\includegraphics{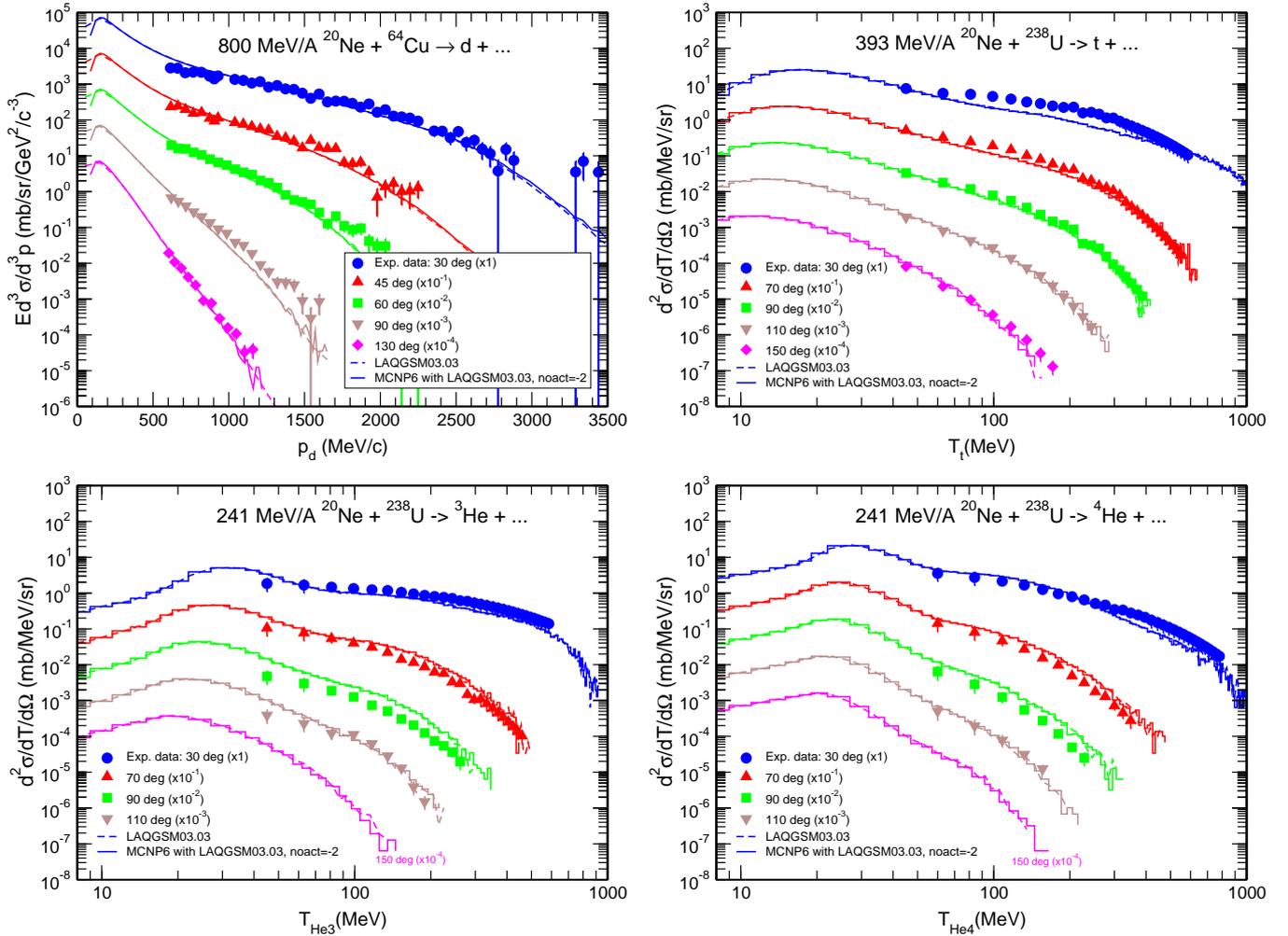}
}
\caption{Experimental \cite{14,22} double-differential spectra of deuterons, 
tritons, $^3$He, and $^4$He at five different angles (symbols) from 
thin $^{64}$Cu and $^{238}$U targets bombarded with $^{20}$Ne beams 
of listed energies compared with the results by LAQGSM03.03 used as a 
stand-alone code and by MCNP6 using the LAQGSM03.03 event-generator,
as indicated (no $^3$He and $^4$He measured data are available at 
150$^\circ$, so we present for this angle only our predictions).}
\end{figure*}

\subsection{Some physics problems and bugs fixes}
\label{sec:2.3}

Naturally, during our extensive V\&V work of MCNP6, we discovered 
some bugs and more serious physics problems in MCNP6 or/and in 
MCNPX.  Most of them have been fixed; we continue our work to 
solve all the observed problems before MCNP6 is distributed to 
the public. Let us present below only two examples. Fig. 8 shows 
the experimental \cite{28} spectra of neutrons at 5$^\circ$, 10$^\circ$, 
20$^\circ$, 30$^\circ$, 40$^\circ$, 60$^\circ$, and 80$^\circ$ from 
a relatively thin Cu target bombarded with a 600 MeV/nucleon $^{28}$Si 
beam compared with results by MCNP6 using LAQGSM03.03 
used as a stand alone code, as well as with calculations by 
MCNPX 2.7.B \cite{29} using LAQGSM03.01. Dr. Igor Remec of ORNL, who 
kindly sent us these (and many other) MCNPX 2.7.B results called 
our attention to a problem he observed in the MCNPX 2.7.B for neutron 
spectra at forward angles. For unknown reasons, MCNPX 2.7.B using 
LAQGSM03.01 strongly overestimates the neutron spectra at forward 
angles (see the cyan lines in Fig. 8), while LAQGSM03.01 used 
as a stand alone code describes such spectra very well (see the 
black lines in Fig. 8). A special investigation by Dr. Mike James 
of the LANL D-5 Group has identified a previously unobserved error 
in the implementation of LAQGSM03.01 in MCNPX, which caused that 
problem. This implementation error was fixed by Mike James in a
very recent version of MCNPX by replacing completely the relatively 
old LAQGSM03.01 with the latest version LAQGSM03.03. Such a 
replacement was also done in MCNP6 by Dick Prael. As we can see 
from Fig. 8, the current version of MCNP6 describes these measured 
neutron spectra very well, just as LAQGSM03.03 and LAQGSM03.01 do 
as stand alone codes.

%
\begin{figure*}                                                       
\vspace*{2mm}       
\centering
\resizebox{0.65\hsize}{!}{
\includegraphics{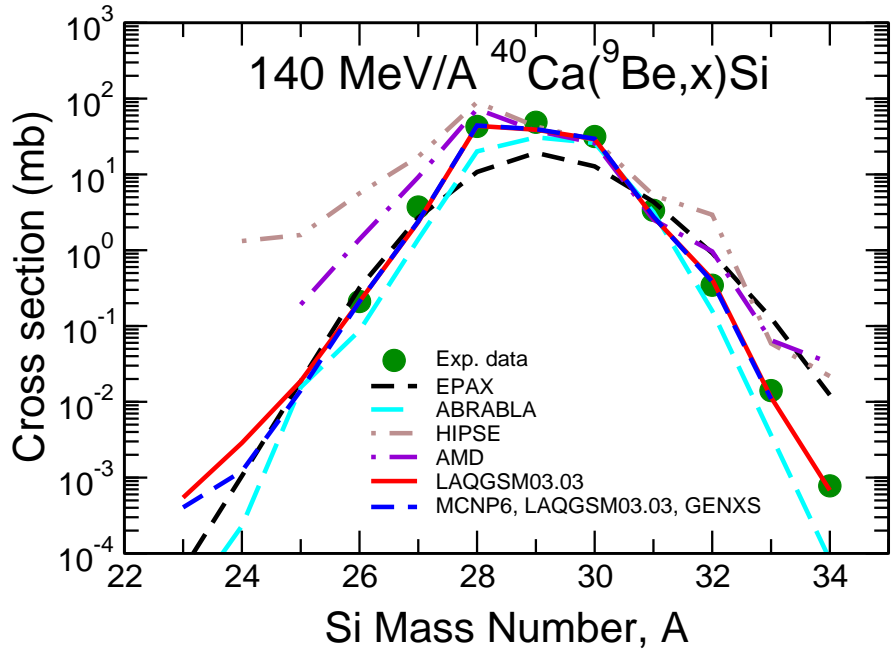}
}
\vspace*{5mm} 
\caption{Experimental \cite{23} mass number distribution of the Si ions 
yield (green filled circles) from the 140 MeV/A $^{40}$Ca + $^9$Be 
reaction compared with results by EPAX \cite{24}, ABRABLA \cite{25}, 
HIPSE \cite{26}, and AMD \cite{27} from \cite{23}, as well as with predictions 
by LAQGSM03.03 used as a stand alone code and by MCNP6  using 
the LAQGSM03.03 event-generator, as indicated.}
\end{figure*}

%
\begin{figure*}                                                       
\vspace*{3mm}       
\centering
\resizebox{0.45\hsize}{!}{
\includegraphics{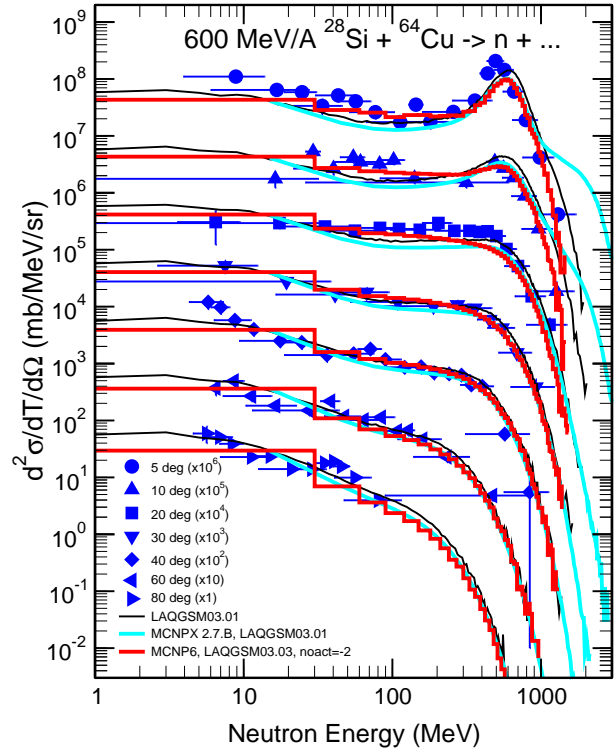}
}
\vspace*{4mm} 
\caption{Experimental \cite{28} neutron spectra at 5$^\circ$, 10$^\circ$, 
20$^\circ$, 30$^\circ$, 40$^\circ$, 60$^\circ$, and 80$^\circ$ from 
the 600 MeV/A $^{28}$Si + Cu reaction compared with results by 
LAQGSM03.01 used as a stand alone code, and with our MCNP6 results 
using LAQGSM03.03, as well as with results by MCNPX 2.7.B \cite{29} 
obtained by Dr. Igor Remec at ORNL using LAQGSM03.01, as indicated.}
\end{figure*}

%
\begin{figure*}                                                       
\centering
\resizebox{1.00\hsize}{!}{
\includegraphics{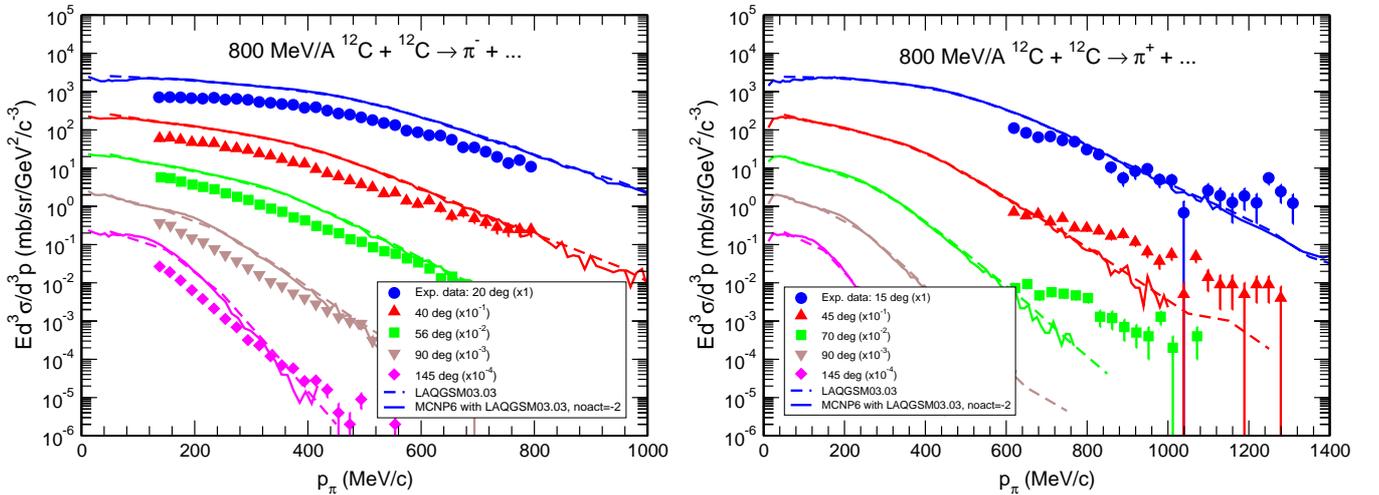}
}
\caption{Experimental \cite{22} invariant $\pi^-$ and $\pi^+$ spectra at five 
different angles (symbols) from a thin C target bombarded with an 800 
MeV/nucleon $^{12}$C beam compared with results LAQGSM03.03 used as a 
stand alone code and by MCNP6 using the LAQGSM03.03 event-generator, 
as indicated (no $\pi^+$ measured data are available at 90 and 145 
degrees, so we present for this angle only our predictions).}
\end{figure*}

Fig. 9. shows an example of another problem solved recently in MCNP6. 
For historical reasons, the initial version of MCNP6 would ``tally" 
(i.e., would count, and provide results in the output file) particles 
together with antiparticles from any nuclear reactions, as would MCNPX 
and MCNP5. This was needed and very useful for some of our applications. 
But for other applications, especially at high-energies, this feature 
is nor convenient at all, as it does not allow users to easily get 
separate results for particles and antiparticles. This issue was 
corrected recently in MCNP6 by Dr. Grady Hughes (in the latest 
versions of MCNPX, MCNPX 2.7.D and MCNPX 2.7.E, this problem was also 
addressed, but in a different way). As we see from Fig. 9, the current version 
of MCNP6 is able to calculate without problems separate spectra of 
$\pi^+$ and $\pi^-$ from nuclear reactions. In this particular example, 
we compare the experimental \cite{22} spectra of $\pi^+$ and $\pi^-$ from 
a thin C target bombarded with a 800 MeV/nucleon $^{12}$C with results 
by MCNP6 using LAQGSM03.03 and with calculations by LAQGSM03.03 used 
as a stand alone code. We see a very good agreement between results by 
MCNP6 and by LAQGSM03.03 used as a stand alone code, and a worse,
especially for $\pi^+$, but still a reasonable agreement with the 
experimental data.

\subsection{Reactions at ultrarelativistic energies}
\label{sec:2.4}

Finally, let us present in Figs. 10 and 11 two examples of reactions 
at ultrarelativistic energies. Fig. 10 shows a test of the capability 
of MCNP6 to describe the same (almost) heavy-ion induced reaction in a 
very large range of incident energies, namely it provides a comparison 
of the experimental \cite{30}-\cite{33} 
charge distributions of product yields from 
559 MeV/nucleon $^{197}$Au + Cu, 10.6 GeV/nucleon $^{197}$Au + Cu, 
and from a very similar but at an ultrarelativistic energy of 158 
GeV/nucleon $^{208}$Pb + Cu reaction compared with results by MCNP6  
using the LAQGSM03.03 event-generator and by LAQGSM03.03 used as a 
stand alone code. Such capabilities of MCNP6 are needed for astrophysical 
applications, particularly for problems of propagation of cosmic rays 
through matter. Let us note that the MCNP6 results shown in Fig. 10 
represent cross sections of the products from both the projectile and 
target nuclei, while the LAQGSM03.03 used as stand alone calculated 
only fragmentation products from the bombarding nuclei. This is why we 
see a good agreement between the MCNP6 and LAQGSM03.03 results and 
the measured projectile fragmentation cross sections (i.e., for 
products heavier than Cu), and a much higher MCNP6 yield of products 
lighter than Cu than the one calculated by LAQGSM03.03 from only the 
projectiles.

In the end, we like to present an example of our work still in 
progress. Namely, Fig. 11 shows a comparison of 
experimental \cite{34,35} invariant spectra of $K^+$ and $t$ from the 
ultrarelativistic reaction 400 GeV p + $^{181}$Ta compared with 
our preliminary results by MCNP6 using LAQGSM03.03 and with 
calculations by LAQGSM03.03 used as a stand alone code, as well 
as with our 2005 results by LAQGSM03.01 from Ref. \cite{36}. Similar 
preliminary results are obtained also for the measured $p$, $d$, 
$^3$He, $^4$He, $K^-$, and antiproton spectra from $^{181}$Ta, as 
well as for all measured spectra from C, Al, and Cu (see \cite{34,35} 
and references therein and our LAQGSM03.01 results presented in 
Refs. \cite{36,37}). In addition to astrophysical applications, such 
reactions are of great academic interest in studying the production 
of so called ``cumulative particles," i.e., energetic particles at 
backward angles in the kinematic region forbidden in interactions 
of the projectile with free stationary nucleons of the target 
nucleus. It is believed that cumulative particles contain information 
needed for the study of the high momentum component of nuclear wave 
functions, or of collective phenomena in nuclei, or of quark and 
gluon degrees of freedom. We see that our preliminary results by 
MCNP6 differ a little from the results obtained with the LAQGSM03.03 
used as a stand alone code, requiring more work on MCNP6. But the 
agreement with the experimental data is reasonably good, especially 
when considering that to the best of our knowledge, these 
``cumulative" particle spectra measured at the Fermi National 
Accelerator Laboratory three decades ago were described simultaneously 
within a single approach and without using any ``exotic" 
nuclear-reaction mechanisms for the first time in 2005 by our 
LAQGSM03.01 \cite{36,37} and here, with MCNP6 using LAQGSM03.03.

%
\begin{figure*}                                                       
\centering
\resizebox{1.00\hsize}{!}{
\includegraphics{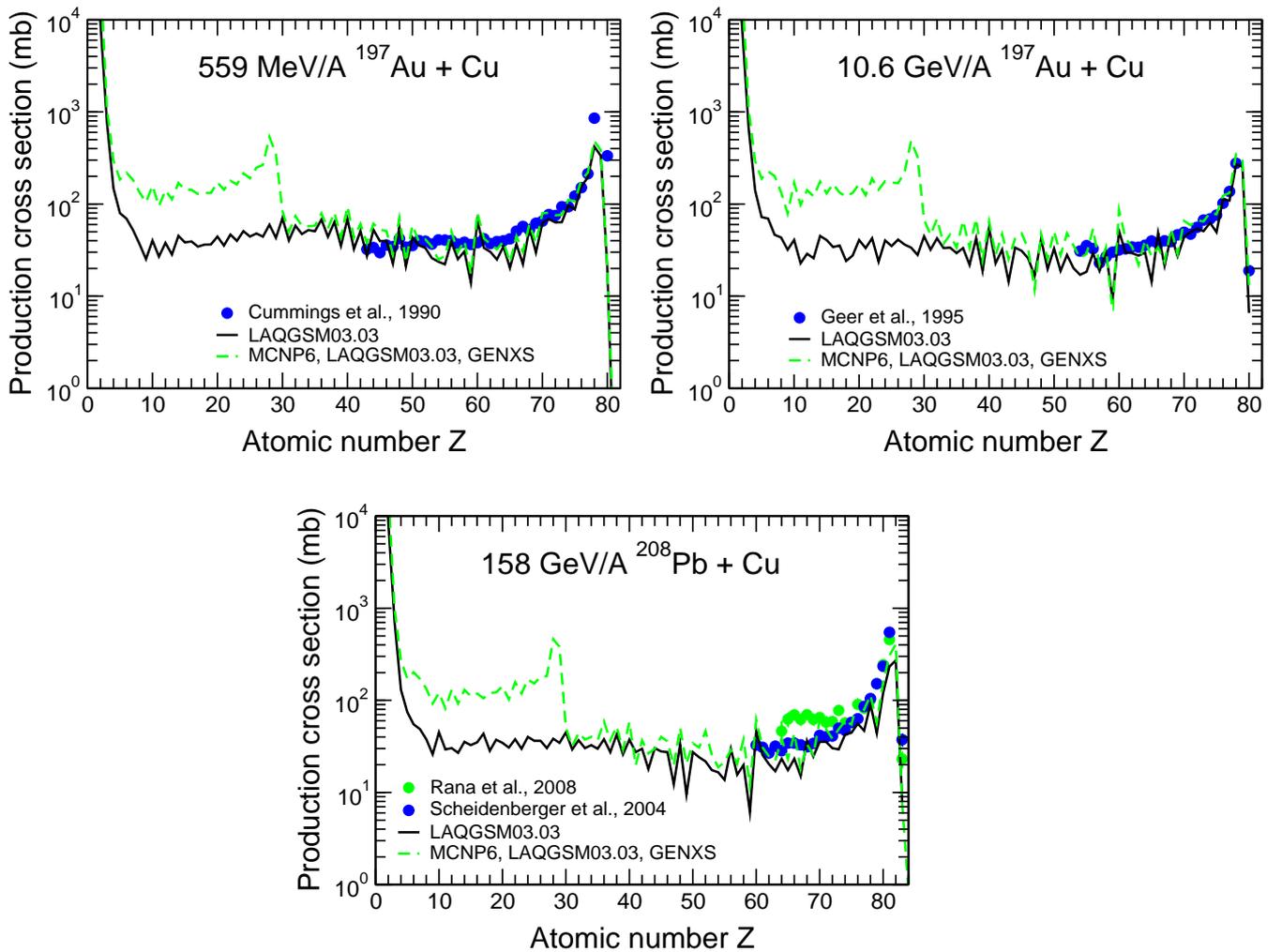}
}
\vspace*{3mm}
\caption{Experimental \cite{30}-\cite{33} charge distributions of product yields 
(color filled circles) from 559 MeV/A $^{197}$Au + Cu, 10.6 GeV/A 
$^{197}$Au + Cu, and 158 GeV/A $^{208}$Pb + Cu reactions compared with 
results by LAQGSM03.03 used as a stand alone code and by MCNP6 sing 
the LAQGSM03.03 event-generator, as indicated.}
\end{figure*}

Let us mention here that the results presented in Fig. 11 correspond
to the status of MCNP6 as of November 2010, when the first draft of the
current work was written for presentation at the M\&C2011 conference 
\cite{38} (this is why we have labeled those results
in legend of Fig. 11 as ``old MCNP6'').
Thereafter, a bug was found in the high-energy portion of MCNP6
and was fixed by Dr. Dick Prael. As can be seen from the upper row
of Fig. 12, the
current version of MCNP6 provides spectra 
(we label in legend of Fig. 12 the results by the latest version of MCNP6
as ``new MCNP6'')
for these reactions and is in
good agreement with results obtained 
when
LAQGSM03.03 is used as a 
stand alone code.
The lower row of Fig. 12 is related to the problem
of MCNP6 tallying particles separately  from antiparticles,
as discussed in Sec. 2.3. We see that the current version of MCNP6,
where this problem was solved by Dr. Grady Hughes as discussed above, 
is able to describe spectra of $\pi^+$ separately from
$\pi^-$; that was not possible in the initial version of MCNP6
and in early versions of MCNPX. We can see a very good agreement 
of the measured spectra of ``cumulative pions'' from this reaction
with our calculations.

\section{Conclusion}
\label{sec:3}
MCNP6, the latest and most advanced LANL transport code representing 
a recent merger of MCNP5 and MCNPX, has been validated and verified 
against a variety of intermediate and high-energy experimental data 
and against calculations by different versions of MCNPX and results 
by several other codes. In the present work, MCNP6 was tested using 
mainly the latest modifications of the Cascade-Exciton Model (CEM) 
and of the Los Alamos version of the Quark-Gluon String Model (LAQGSM) 
event generators CEM03.03 and LAQGSM03.03. We found that MCNP6 describes 
reasonably well various reactions induced by particles and heavy-ions 
at incident energies from 18 MeV to about 1 TeV/nucleon measured on 
both thin and thick targets and agrees very well with similar results 
obtained with MCNPX and calculations by other codes. Most of several 
computational bugs and more serious physics problems observed in 
MCNP6/X during our V\&V have been fixed. We continue our work to solve 
all the known problems before the official distribution of MCNP6 to 
the public via RSICC at Oak Ridge, TN, USA planned for the middle 
of 2011. From the current V\&V, we can conclude that MCNP6 is a reliable 
and useful transport code for different applications involving reactions 
induced by almost all types of elementary particles and heavy-ions, 
in a very broad range of incident energies.

%
\begin{figure*}                                                       
\centering
\resizebox{1.00\hsize}{!}{
\includegraphics{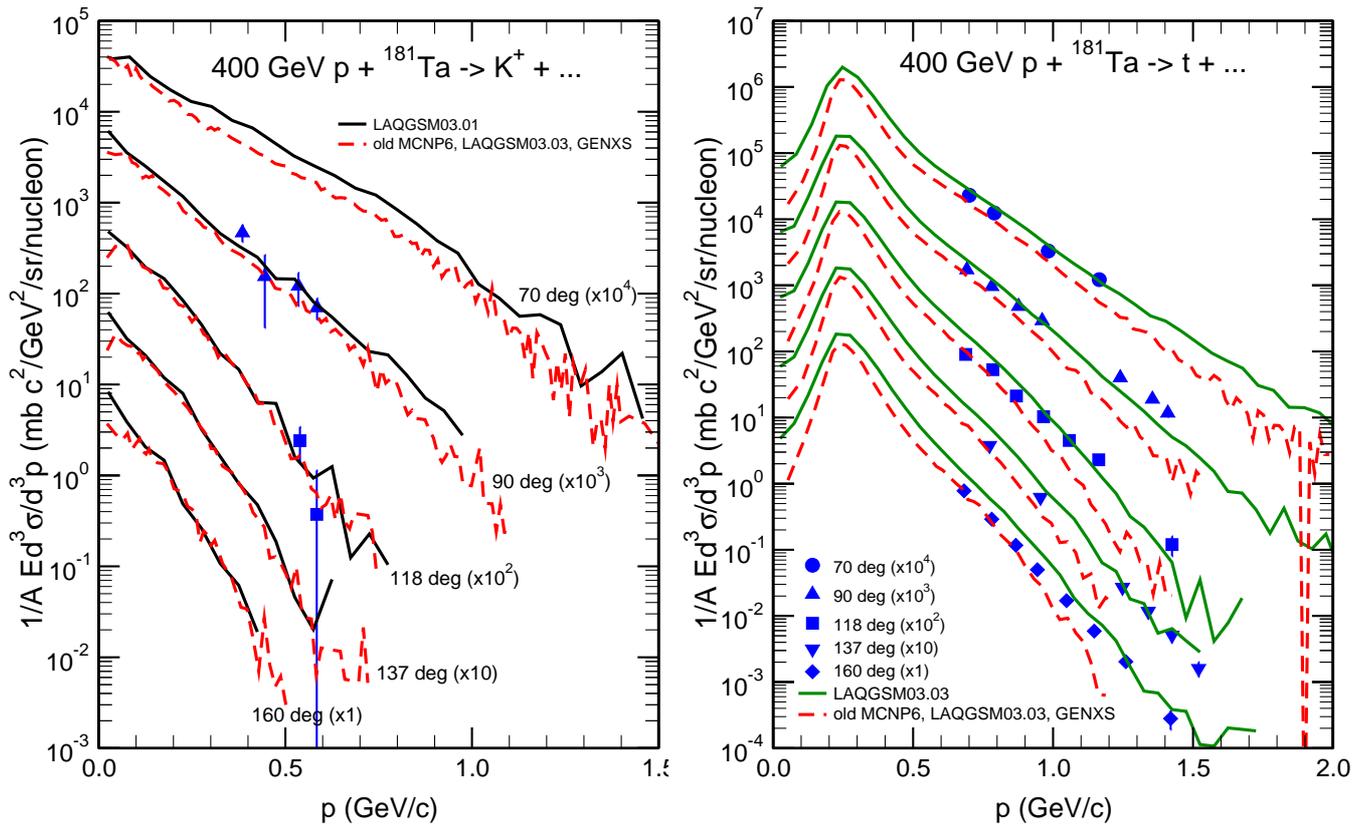}
}
\caption{Experimental \cite{34,35} invariant spectra of $K^+$ and $t$ 
from the reaction 400 GeV p + $^{181}$Ta (symbols) compared with 
results by a preliminary, still under development, version (see text) of 
MCNP6 using LAQGSM03.03 (red dashed lines), with 
calculations by LAQGSM03.03 used as a stand alone code, and with 
our 2005 results by LAQGSM03.01 from Ref. \cite{36}, as indicated. 
Similar results are obtained also for the measured $p$, $d$, $^3$He, 
$^4$He, $K^-$, and antiproton spectra from $^{181}$Ta, as well 
as for all measured spectra from C, Al, and Cu (see \cite{34,35} and 
references therein and our LAQGSM03.01 results in Refs. \cite{36,37}).}
\end{figure*}

%
\begin{figure*}                                                       
\centering
\resizebox{1.00\hsize}{!}{
\includegraphics{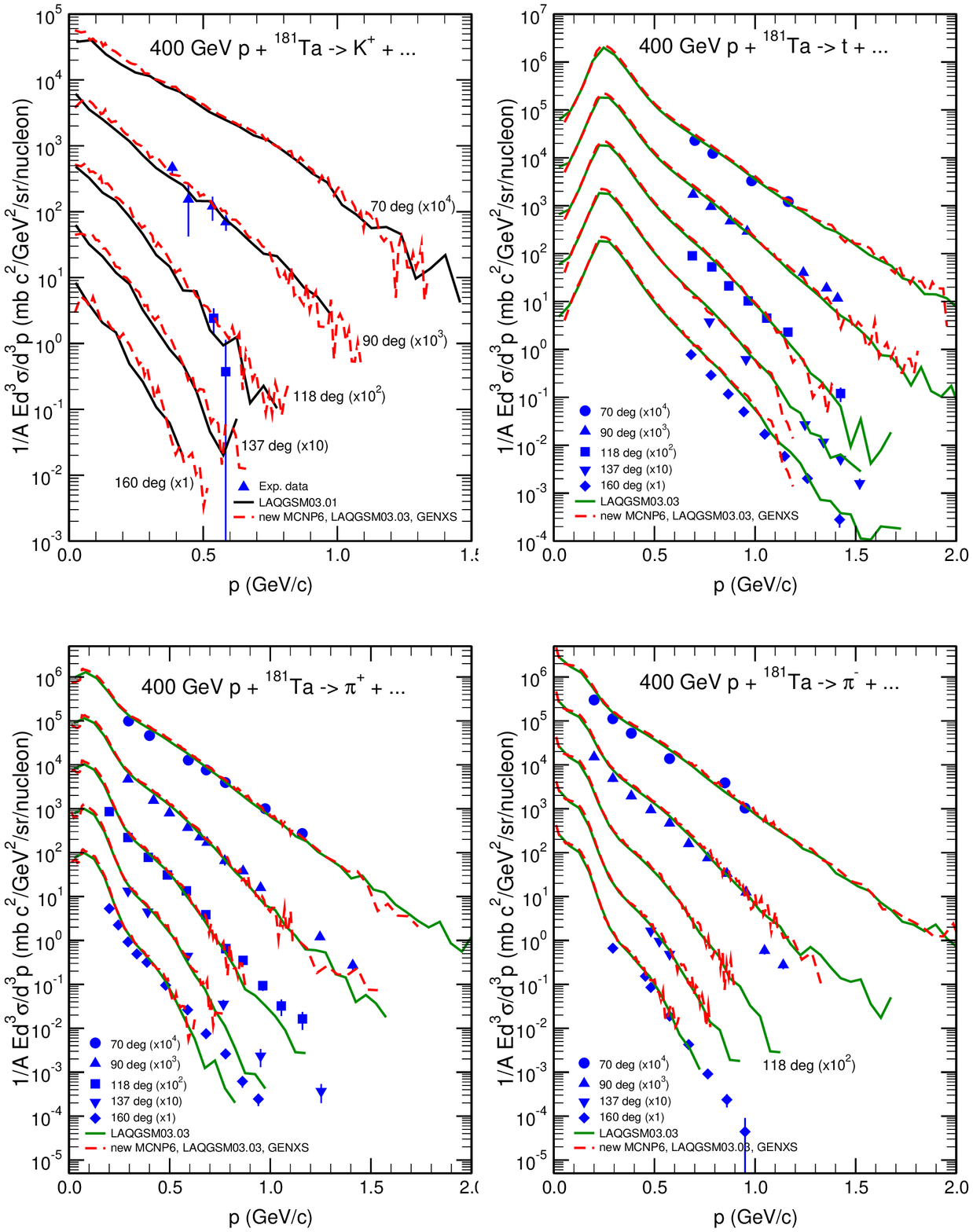}
}
\caption{Experimental \cite{34,35} invariant spectra of $K^+$, $t$, 
$\pi^+$, and $\pi^-$
from the reaction 400 GeV p + $^{181}$Ta (symbols) compared with 
results by the latest version (see text) of 
MCNP6 using LAQGSM03.03 (red dashed lines), with 
calculations by LAQGSM03.03 used as a stand alone code, and with 
our 2005 results by LAQGSM03.01 from Ref. \cite{36}, as indicated.} 
\end{figure*}

\begin{acknowledgement}
I am grateful to my LANL colleagues, Drs. Forrest Brown, Jeff Bull, 
Tim Goorley, Grady Hughes, Mike James, Roger Martz, Dick Prael, 
Arnold Sierk, and Laurie Waters as well as to Franz Gallmeier and 
Wei Lu of ORNL, Oak Ridge, TN, USA, and to Konstantin Gudima of 
the Academy of Science of Moldova, for useful discussions, help, 
and/or for correcting some of the MCNP6/X bugs I have detected 
during the current V\&V work.

I thank Drs. Vladimir Belyakov-Bodin, Masayuki Hagiwara, Yasuo Miake,  
Shoji Nagamiya, Paolo Napolitani, Meiring Nortier, Yusuke Uozumi, 
and John Weidner for sending me their publications or/and files with 
numerical values of their experimental data I used in my V\&V work. 
It is a pleasure to acknowledge Drs. Franz Gallmeier, Masayuki 
Hagiwara, Antonin Krasa, Wei Lu, Mitja Majerle, and Igor Remec 
for sending me their MCNPX calculations included in my comparisons.

This work was carried out under the auspices of the National Nuclear 
Security Administration of the U.S. Department of Energy at Los Alamos 
National Laboratory under Contract No. DE-AC52-06NA25396 with funding 
from the Defense Threat Reduction Agency (DTRA).

\end{acknowledgement}

%
%

\end{document}